\documentclass[11pt,twoside
]{article}
\usepackage{baaa2008}
\usepackage{graphicx}
\usepackage{subfigure}
\usepackage{psfrag}
\usepackage{amssymb}
\usepackage[spanish,activeacute,english]{babel}
\usepackage[latin1]{inputenc}
\usepackage[T1]{fontenc} 
\usepackage{ae,aecompl} 
\usepackage{latexsym}
\usepackage{verbatim}
\usepackage{amsmath}
\usepackage{amsfonts}
\usepackage{amssymb}
\usepackage{wasysym}
\usepackage[colorlinks=true,dvips]{hyperref}

\begin{document}
\myselectenglish
\vskip 1.0cm
\markboth{N. Jim\'enez et al.}%
{The Color-Magnitude Relation of cluster galaxies}

\pagestyle{myheadings}
\vspace*{0.5cm}
\parindent 0pt{PRESENTACI\'ON MURAL}
\vskip 0.3cm
\title{The Color-Magnitude Relation of Cluster Galaxies: Observations and Model Predictions}
\author{N. Jim\'enez$^{1,2}$, A. Smith Castelli$^{1,2}$,  S. A. Cora$^{1,2}$ \& L. P. Bassino $^{1,2}$}
\affil{
(1) Facultad de Ciencias Astron\'omicas y Geof\'isicas (FCAG, UNLP)\\
(2) Instituto de Astrof\'isica de La Plata (CCT La Plata, CONICET - UNLP) and CONICET, Argentina\\
}
\begin{abstract} We investigate the origin of the color-magnitude relation 
(CMR) observed in cluster galaxies by using a combination of cosmological 
{\em N}-body/SPH simulations of 
galaxy clusters, 
and a semi-analaytic model 
of galaxy 
formation (Lagos, Cora \& Padilla 2008). Simulated results are compared with 
the photometric properties of early-type galaxies in the Antlia cluster
(Smith Castelli et al. 2008). The good agreement obtained between
observations and simulations allows us to use the information provided by the 
model for unveiling the physical processes that yield the tigh observed CMR.
\end{abstract}
\begin{resumen}
Investigamos la relaci\'on color-magnitud (CMR) observada en c\'umulos
de galaxias, usando simulaciones hidrodin'amicas cosmol\'ogicas de
{\em N}-cuerpos 
de c\'umulos
de galaxias junto con un modelo semianal\'itico de formaci\'on de
galaxias (Lagos, Cora \& Padilla 2008). Los resultados de las
simulaciones son comparados con las propiedades fotom\'etricas de galaxias
de tipo temprano del  c\'umulo de Antlia (Smith Castelli et al.
2008). El buen acuerdo obtenido entre las observaciones y los
resultados del modelo, nos permiten utilizar la informaci\'on
suministrada por \'este en el estudio de los procesos f\'isicos que
conducen a una CMR observada muy bien definida.
\end{resumen}

\section{Introduction}
Early-type galaxies residing in groups and cluster of galaxies
define a sequence
in the colour-magnitude diagram, being bright galaxies 
redder than fainter ones.
This colour-magnitude relation (CMR)
seems to be universal in nearby clusters of galaxies. 
Smith Castelli et al. (2008, hereafter SC08) have recently obtained a 
linear fit to the CMR of
the Antlia cluster with a slope in agreement with those found in Virgo
(Lisker et al. 2008), Fornax (Mieske et al. 2007),
Perseus (Conselice et al. 2002) and Coma (L\'opez-Cruz et al. 2004). Such
universality lead several authors to 
suggest that the build up of this
relation in galaxy clusters is more related to galaxies internal processes
than to the influence of the environment (SC08; 
Misgeld et al. 2008).

The understanding of the building of the CMR displayed by elliptical galaxies
is a key test for galaxy formation models. 
Using hydrodynamical simulations of
groups and clusters of galaxies, Saro et al (2006) and Romeo et al. (2008) 
have tried to reproduce 
the observed slope and normalization of the CMR.
Semi-analytic models have also been used for this kind of study
(De Lucia et al. 2004: Kaviraj et al. 2005).
None of these galaxy formation models considers the effect of
feedback from active glactic nuclei (AGN), 
which is essencial to avoid the formation
of too massive and blue cluster dominant galaxies.

We present a study on the origin of CMR 
in galaxy clusters by applying a semi-analytic model of galaxy formation to
the outputs of hydrodynamical non-radiative 
{\em N}-body/SPH 
numerical simulations of clusters of galaxies.
We compare the results obtained from this model with
the galaxy properties of the Antlia cluster (SC08).
The Antlia cluster is the third nearest well populated galaxy
cluster after Virgo and Fornax ($D=35.2$ Mpc). 

\vskip 0.5cm

\section{The Model}

We use
a combination of cosmological adiabatic {\em N}-body/SPH simulations of 
clusters of galaxies
and the SAG (acronym for `Semi-Analytic Galaxies') semi-analytic model of
galaxy formation (Lagos et al. 2008). 
This model follows the formation
and evolution of galaxies including gas cooling, star formation,
feedback from supernovae explosion and galaxy mergers, a detailed
implementation of the metal enrichment of the baryonic component, and 
feedback from AGN.

We consider two simulated galaxy clusters, having virial masses in the range 
$\simeq (1-13)\times 10^{14}\,h^{-1}\,{\rm M}_\odot$ (Dolag et al. 2005).  
These clusters  
have been initially selected from a $\Lambda$ cold dark matter simulation 
of a cosmological
box of $479\,h^{-1}$~Mpc of size, characterized by 
$\Omega_{\rm m}$=0.3, $\Omega_{\Lambda}$=0.7, 
${\rm H}_{\rm o}= 70 \, {\rm km} \, {\rm s}^{-1} \, {\rm Mpc}^{-1}$,
$\Omega_{\rm b}=0.039$ for the baryon density parameter, and 
$\sigma_8=0.9$ for the normalization of the power spectrum.
The mass resolution is 
$m_{\rm dm}=1.13 \times 10^{9}\,h^{-1}\,{\rm M}_{\odot}$ and 
$m_{\rm gas}=1.69\times 10^{8}\,h^{-1}\,{\rm M}_{\odot}$, 
for dark matter and gas particles, respectively. 

\section{Color-Magnitude and Luminosity-Metallicity Relations}

In order to compare simulations with observations, we
apply a morphological criterium to select elliptical
galaxies from the model. 
Ellipticals are those bulge dominated systems
where the ratio between the bulge mass and the
total stellar mass, $r= M_{\rm Bulge}/M_{\star}$,
satisfy the condition $r> 0.95$. 

X-ray observations have revealed that the Antlia has an average temperature
of $kT \simeq 2.0$ keV (Pedersen et al. 1997, Nakazawa et al. 2000).
The virial temperature of the least massive cluster considered 
($kT \sim 1.3$~KeV)
is quite similar to that of
the Antlia cluster. 
However, since the general trends of the results are
similar for both simulated clusters, we show here the CMR and
luminosity-metallicity relation for the more massive one, which contains
a larger population of galaxies.

\begin{figure}[h]
  \centering
    \includegraphics[width=.50\textwidth]{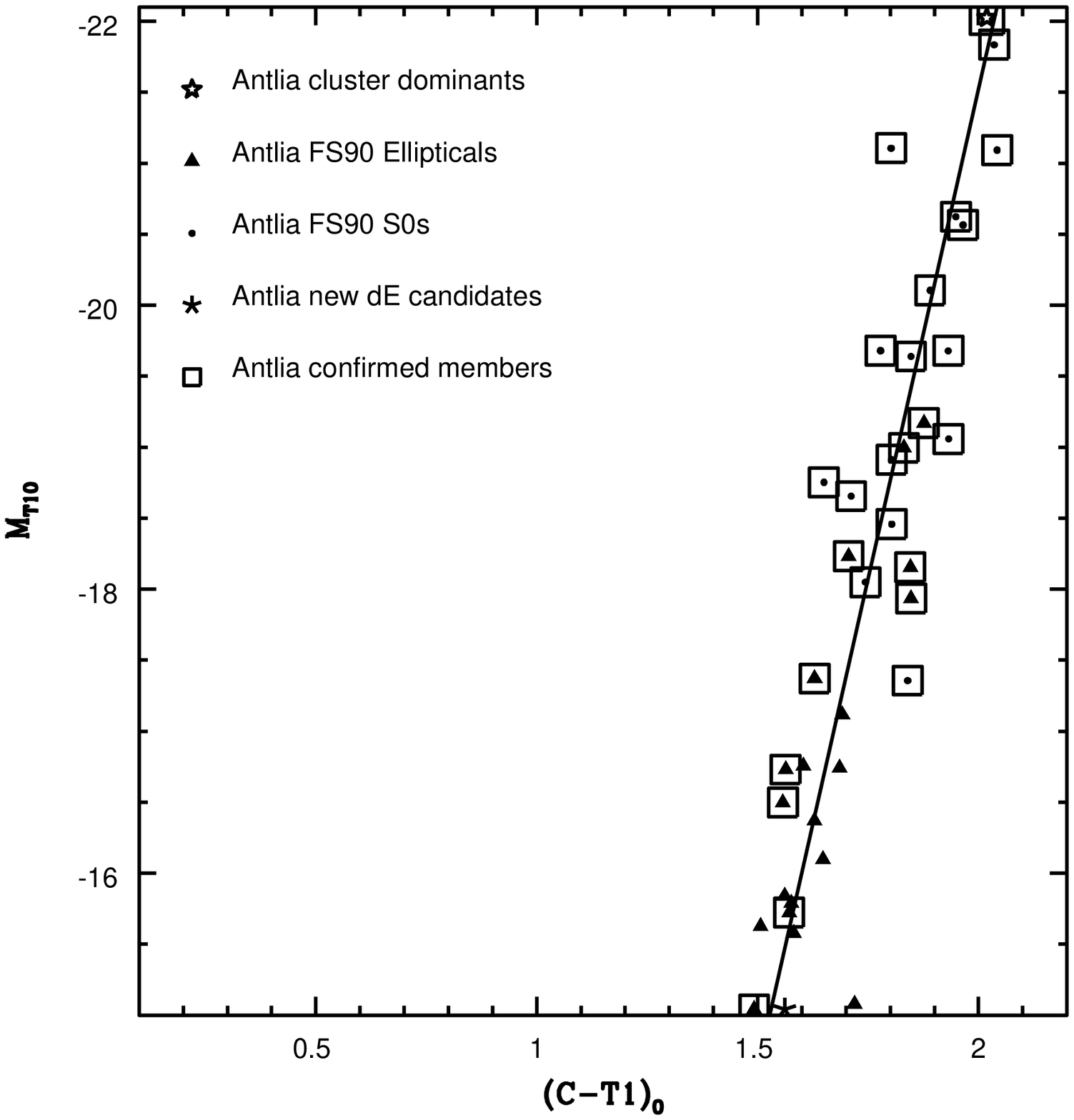}~\hfill%
  \includegraphics[width=.50\textwidth]{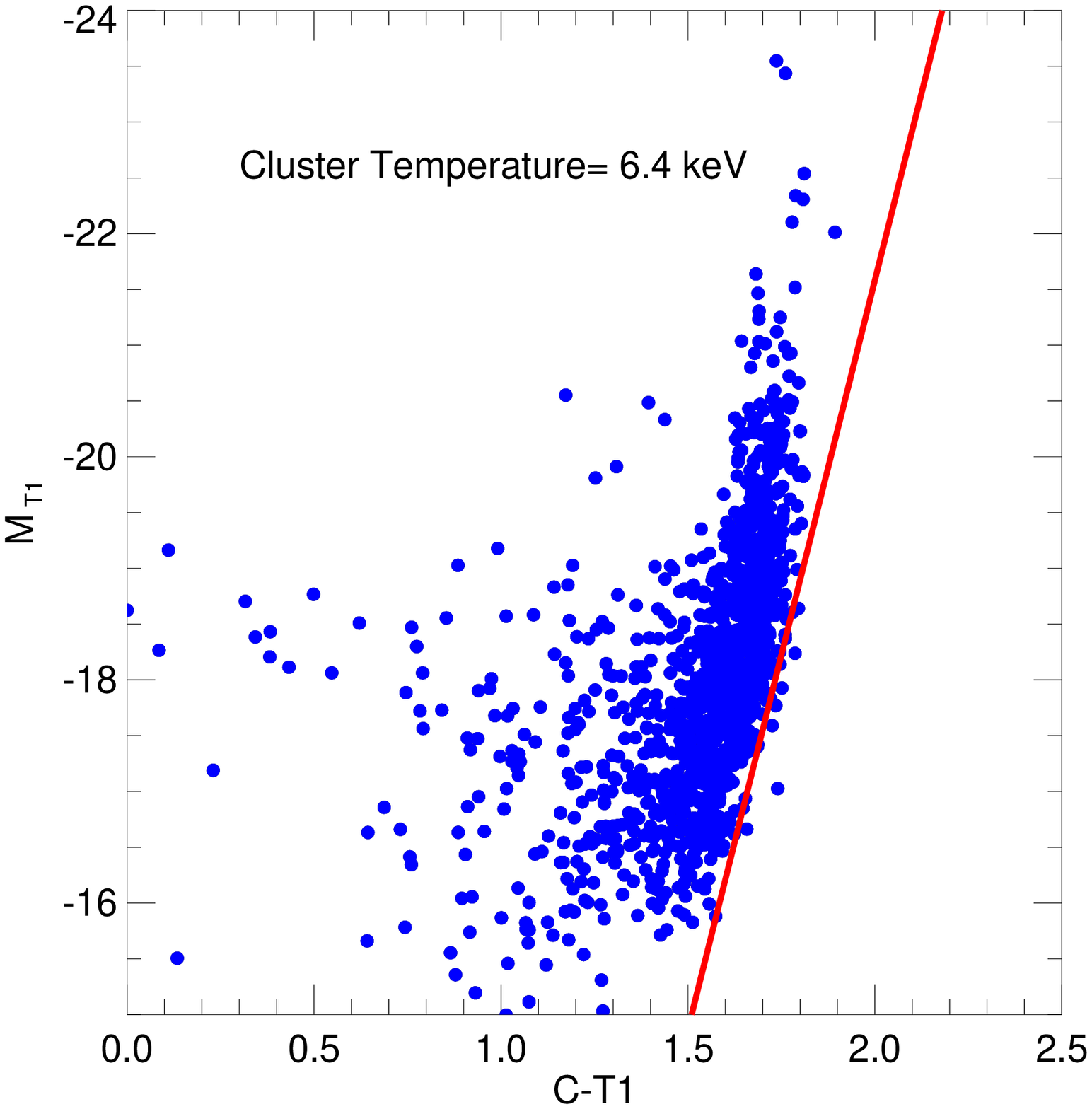}~\hfill
  \caption{{\it Left:} Observed CMR for the Antlia cluster. 
{\it Right:} Simulated CMR 
compared to the mean observed CMR
(solid line).}
  \label{fig:ab1}
\end{figure}

The left panel of figure \ref{fig:ab1} shows the CMR of early-type galaxies
of  the Antlia cluster obtained by
SC08 from CCD wide-field (MOSAIC-CTIO) photometry in
the Washington photometric system ($T_1$ and $C$ filters). This relation is
defined by 51 early-type galaxies from the Ferguson \& Sandage (1990) Antlia
Group catalogue and 21 new early-type dwarf galaxy candidates and members.
The CMR spans 11 mag in brightness
with no change of slope. 

The semi-analytic model provides galactic magnitudes in the Johnson
photometric system. They were converted to the Washington one through 
the transformations given by Forbes
\& Forte (2001) for globular clusters, assuming that early-type
galaxies are old stellar systems. 
Additional conversions
were obtained from Fukugita et al. (1995).
The right panel of figure \ref{fig:ab1} shows the simulated photometric 
properties of early-type galaxies compared to the mean CMR of 
early-type members of  the Antlia cluster. 
The slope denoted by
the red side of the locus occupied by the simulated galaxies
is in very good agreement with the mean observed CMR.

\begin{figure}[!ht]
  \centering
  \hfill\includegraphics[width=.50\textwidth]{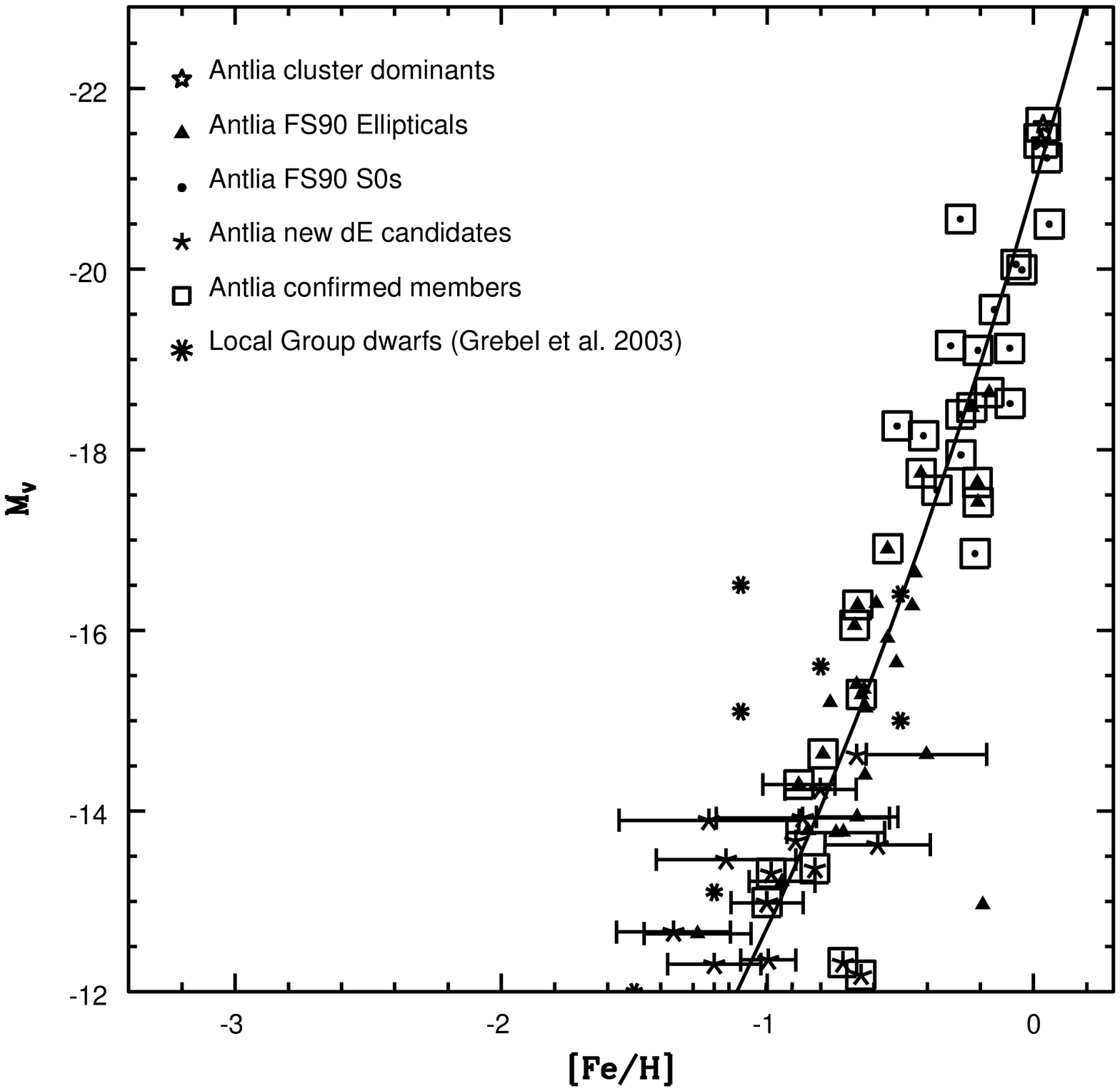}~\hfill%
  \includegraphics[width=.50\textwidth]{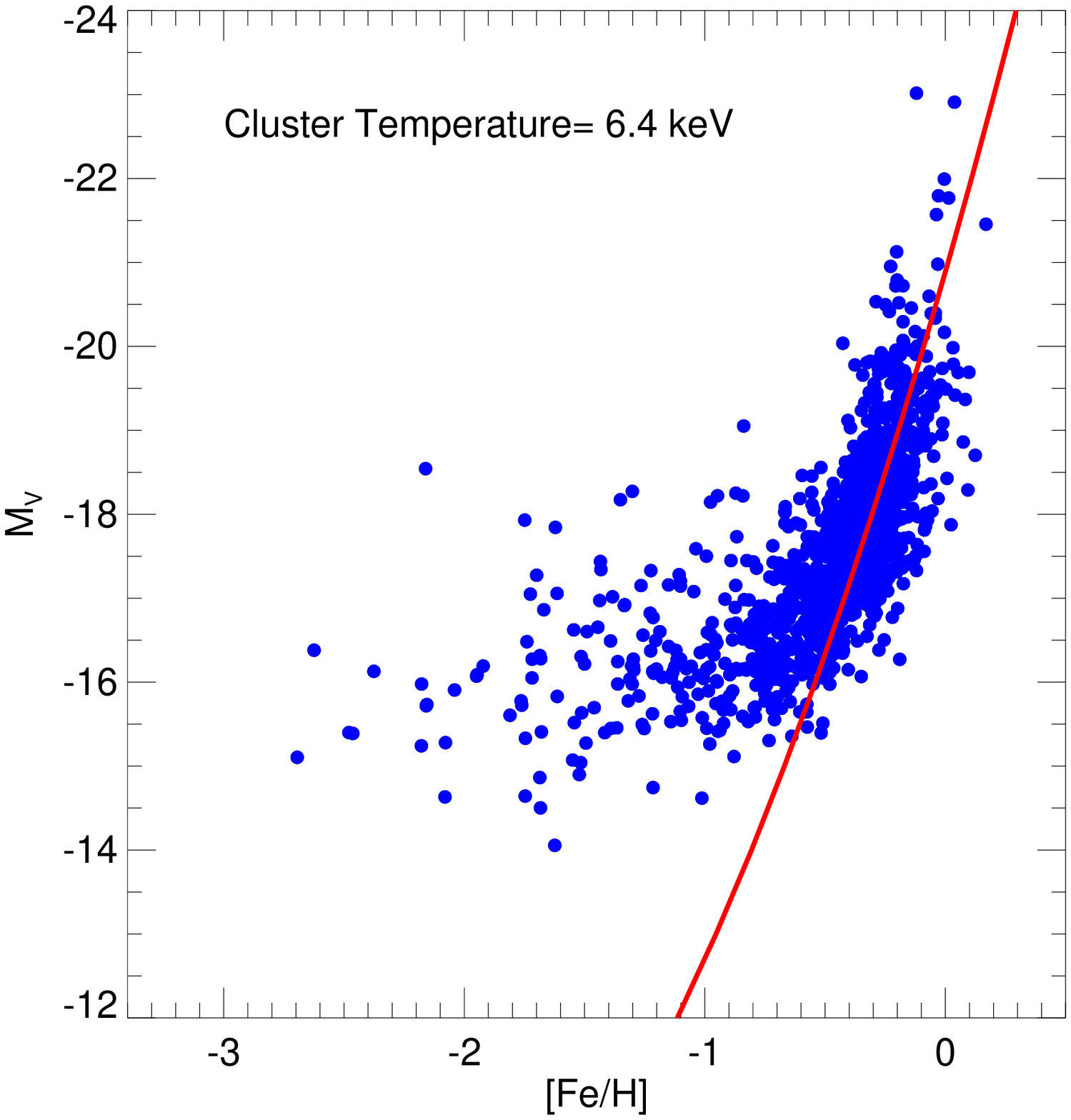}\hfill~
  \caption{{\it Left:} Observed luminosity-metallicity relation  
for the Antlia cluster.
{\it Right:} The same relation for the simulated 
cluster 
compared with the mean observed relation for Antlia 
(solid line).}
 \label{fig:ab2}
\end{figure}

The observed $M_{V}$ magnitude versus $[{\rm Fe}/{\rm H}]$ relation
for the Antlia
galaxies (SC08) and Local Group dwarfs (Grebel
et al. 2003) is shown in the left panel of figure \ref{fig:ab2}. 
Antlia galaxies 
metallicities were obtained by transforming $(C-T_{1})$ colors to 
$[{\rm Fe}/{\rm H}]$ 
values through the Harris \& Harris (2002) relation for globular clusters.
The right panel of figure \ref{fig:ab2}
shows the
corresponding relation obtained from the larger simulated cluster which is
compared to the mean observed relation. We find an excellent
agreement for $[{\rm Fe}/{\rm H}] > -1$. 
The spread of lower luminous objects towards
lower metallicities is due to the presence of late-type galaxies,
missclasiffied as ellipticals by the rather uncertain threshold in
the adopted morphological criterium. This set of simulated galaxies
also populate the blue side of the color-magnitude diagram.

The similar trends found in the colour-magnitude and metallicity-luminosity 
relations between observations and simulations are
encouraging. We plan to extend this study in order to explain
the physical origin of the dispersion of the observed CMR, thus
evaluating the influence of the star formation history and the
chemical enrichment of the involved galaxies.

\acknowledgements
This work was supported by 
grants from Consejo Nacional de Investigaciones Cient\'ificas y T\'ecnicas
(CONICET), 
Agencia Nacional
de Promoci\'on Cient\'ifica Tecnol\'ogica and Universidad Nacional de
La Plata, Argentina.

\begin{referencias}
\vskip 0.5cm

\reference Conselice, C.J., Gallagher, J.S. III \& Wyse, R.F.G. 2002, \aj, 123, 2246 
\reference De Lucia, G., Poggianti, B.M., Arag\'on-Salamanca, A., et al. 2004, \apj, 610, L77  
\reference Dolag, K., Vazza, F., Brunetti, G. \& Tormen, G.G. 2005, \mnras, 364, 753 
\reference Ferguson, H.C. \& Sandage, A. 1990, \aj, 100, 1
\reference Forbes, D.A., Forte, J.C. 2001, \mnras, 322, 257 
\reference Fukugita, M., Shimasaku, K. \& Ichikawa, T. 1995, \pasp, 107, 945
\reference Grebel, E.K., Gallagher, J.S. \& Harbeck, D. 2003, \aj, 125, 1926 
\reference Harris, W.E. \& Harris, G.L.H. 2002, \aj, 123, 3108 
\reference Kaviraj, S., Devriendt, J. E. G., Ferreras, I. \& Yi, S. K. 2005, \mnras, 360, 60  
\reference Lagos, C., Cora, S.A. \& Padilla, N.D. 2008, \mnras, 388, 587 
\reference Lisker, T., Grebel, E.K. \& Binggeli, B. 2008, \aj, 135, 380
\reference L\'opez-Cruz, O., Barkhouse, W.A. \& Yee H.K.C. 2004, \apj, 614, 679 
\reference Mieske, S., Hilker, M., Infante, L. \& Mendes de Oliveira, C. 2007, A\&A, 463, 503 
\reference Misgeld, I., Mieske, S. \& Hilker, M. 2008, A\&A, 486, 697 
\reference Nakazawa, K, Makishima, K., Fukazawa, Y. \& Tamura T. 2000, \pasj, 52, 623
\reference Pedersen, K., Yoshii, Y. \& Sommer-Larsen, J. 1997, \apj, 485, L17
\reference Romeo, A.D., Napolitano, N.R., Covone, G., et al. 
2008, \mnras, 389, 13  
\reference Saro, A., Borgani, S., Tornatore, L.,  
et al. 2006, \mnras, 373, 397
\reference Smith Castelli, A., Bassino, L., Richtler, T., et al. 2008, MNRAS, 386, 2311 (SC08) 
\end{referencias}

\end{document}